\documentclass[%
 reprint,
 amsmath,amssymb,
 aps,
onecolumn,
]{revtex4-2}

\usepackage{graphicx}% Include figure files
\usepackage{dcolumn}% Align table columns on decimal point
\usepackage{bm}% bold math
\usepackage{xcolor}
\begin{document}

\preprint{APS/123-QED}

\title{Contribution of spanwise and cross-span vortices to the lift generation of low-aspect-ratio wings: Insights from force partitioning}% Force line breaks with \\

\author{Karthik Menon}
\email{karthikmenon@stanford.edu}
\altaffiliation[Current affiliation: ]{Stanford University, Stanford, CA, USA.}%Lines break automatically or can be forced with \\
\author{Sushrut Kumar}%
\affiliation{%
 Department of Mechanical Engineering, Johns Hopkins University, Baltimore, MD, USA
}%
\author{Rajat Mittal}%
 \email{mittal@jhu.edu}
\affiliation{%
 Department of Mechanical Engineering, Johns Hopkins University, Baltimore, MD, USA
}%

%\date{}% It is always \today, today,

\begin{abstract}
This study reports on the vortex-induced lift production mechanisms in low Reynolds number flows over low aspect-ratio rectangular wings. We use a rigorous force partitioning method which allows for the estimation of the pressure-induced aerodynamic loads due to distinct flow features or vortex structures in the flow around the wing. The specific focus of this work is on distinguishing the effect of spanwise and cross-span oriented vortex structures on pressure-induced lift production. We quantify the lift induced on the wing by these different vortices, and also estimate their influence within different regions of the flow-field around the wing and in the wake. By varying the aspect-ratio and angle-of-attack of the wing, we show that for most cases, the spanwise oriented vorticity contributes less to the total lift than cross-span oriented vortices. Furthermore, the spanwise vorticity in the near wake is capable of producing net negative lift on the wing and this is explained by separating and quantifying the influence of vortex cores and regions of strain in the wake. The results demonstrate the utility of the force partitioning method for dissecting the flow physics of vortex dominated flows.
%Due to the non-linearities associated with vortex stretching and tilting, rotation-dominated and strain-dominated regions of vorticity exhibit different spatio-temporal evolution over the wing and have important effects on the lift production due to spanwise and cross-spanwise vorticity. 
%Finally, we compare these findings for a rectangular and delta wing and show that suppressed vortex shedding in the case of the delta wing changes this wake behavior and has interesting consequences for the influence of wake vortices on lift production. 
\end{abstract}

\maketitle

%\tableofcontents

\section{Introduction}
\label{sec:intro}

Coherent vortex structures are known to play a key role in a large variety of problems in fluid dynamics. In many fundamental problems such as aerodynamic force generation, separation bubbles and laminar to turbulent transition, past work has highlighted the importance of vortex structures which exhibit specific shapes and orientations that have a large influence on the dynamics of the flow. Examples include vortex rings \citep{Drucker1999LocomotorVelocimetry}, spanwise rollers, streamwise streaks, etc. Particularly in the domain of problems relating to aerodynamics and hydrodynamics, the distinct roles played by different kinds of coherent vortex structures, such as streamwise-oriented tip vortices and predominantly spanwise-oriented leading-edge vortices, has been a focus of significant research and has been shown to be of fundamental importance \citep{shyy2007aerodynamics,Eldredge2019}.

Efforts to analyze the dynamics of distinct vortex structures and their interactions have been mainly motivated by the need to understand their influence on aerodynamic load generation. For example, the importance of streamwise-oriented tip-vortices, which have a significant impact on the lift produced by finite-span wings, has been recognized in the earliest models of inviscid flow \citep{kundu2008fluid}. Similarly, the leading-edge vortex (LEV) has been shown to be a major lift-enhancement mechanism in a variety of aerodynamics applications \citep{Ellington1996Leading-edgeFlight,Eldredge2019}. The motion and shedding of LEVs is implicated in phenomena crucial to the aerodynamic loads generated on wings, such as in dynamic stall \citep{Carr1988ProgressStall,Jantzen2014VortexPlates}. This has spurred significant research into the stability and timing of LEV shedding, and how it relates to force production in various applications from insect and bird flight \citep{Dickinson1993UnsteadyNumbers,Sane2003TheFlight} to swimming locomotion \citep{Zhu2002Three-dimensionalSwimming,Dong2006WakeFoils}.

Several previous studies have also developed methods to quantify the aerodynamic loads induced by individual vortices from flow-field data. For instance, \cite{PittFord2013LiftVortex} and \cite{Onoue2016VortexPlate} analyzed the aerodynamic loads induced by the leading and trailing-edge vortices over airfoils undergoing various maneuvers. The common features of these types of approaches are: (a) the assumption of two-dimensionality; (b) use of inviscid models with point vortices in the field combined with conformal mapping; (c) parameterization of vortex circulation from data;  and (d) calculation of induced aerodynamic load via the Blasius formula \citep{BATCHELOR}. The predictions from such models can be quite good and lead to excellent insights into vortex-induced aerodynamic loads. However, the extension of these inviscid-model based methods to three-dimensional flow fields is non-trivial, and would be particularly difficult for low-aspect-ratio wings, where three-dimensional effects are substantial.

For finite-span wings such as the one that is the focus of the current study, three-dimensional vortex structures have been shown to exhibit a highly complex evolution due to their mutual interactions in the vicinity of the wing as well as in the wake \citep{Taira2009Three-dimensionalNumbers,Zhang2020OnEffects}. Due to this, the dynamics of these dominant vortex structures and their influence on aerodynamic loads has been shown to depend on a variety of factors. For instance, while the tip-vortex is generally considered detrimental to lift on static wings, it can be beneficial in flapping flight \citep{Shyy2009CanWing,Trizila2011Low-reynolds-numberPlate}. Similarly, interactions between the LEV and tip-vortices in flows over finite wings effect their influence on the production of aerodynamic forces on such wings \citep{Taira2009Three-dimensionalNumbers,Jantzen2014VortexPlates}. These complexities have motivated numerous studies of the effect of wing shape, kinematics and other factors on the three-dimensional vortex dynamics over such wings as well as their associated aerodynamic load generation \citep{Lentink2009RotationalWings,Shields2012EffectsWings,Carr2013Finite-spanRatio, Harbig2013ReynoldsPlanforms,Zhang2020LaminarWings}

While these past studies have gone a long way in highlighting the important vortex structures and their influence on aerodynamic load generation in flows over finite wings, they have mostly resorted to \emph{qualitatively} correlating the evolution of particular vortices in the flow to the observed aerodynamic loads on the wing. However, a more \emph{quantitative} analysis that directly relates these vortices and their spatio-temporal dynamics to aerodynamic loads would be useful for the aerodynamic design of control surfaces as well as the the development of effective flow control methods that can more precisely target the most important flow features. Furthermore, in recent work \citep{Menon2021a} we showed that although the majority of past work has focused on the role of vortices, i.e. \emph{rotation-dominated} regions of the flow, in aerodynamic force generation, these vortices are in fact surrounded by regions of \emph{strain-dominated} flow which can have a significant influence on aerodynamic forces and whose contributions are often overlooked.

The focus of this work is therefore to address the aforementioned issues by providing new insights into the fundamental physics of lift generation on finite wings and highlight the important vortex dynamics underlying the aerodynamics of such problems. To that end, we demonstrate the use of a force partitioning method (FPM) to \emph{quantitatively} evaluate the distinct contributions of important types of vortex structures on aerodynamic load generation on finite wings. In addition, this method allows us to separate the influence of rotation-dominated regions from strain-dominated regions of the flow in the generation of aerodynamic loads.

The force partitioning method used here is based on the work of Quartappelle and Napolitano \cite{Quartappelle1982ForceFlows}, Wu \citep{Wu1981TheoryFlows} and Howe \citep{Howe1995OnNumbers} with later extensions by Chang \citep{Chang1992PotentialFlow}, Zhang \textit{et al.} \citep{Zhang2015CentripetalInsects}, and Menon and Mittal \citep{Menon2020,Menon2021}. This method allows for the precise estimation of the loads induced on bodies immersed in a fluid flow by different physical mechanisms, such as unsteady and viscous effects, as well as distinct flow structures. In previous work by the present authors \citep{Menon2021}, we combined this force partitioning method with a data-driven framework to track individual vortices in complex two-dimensional flow-fields and quantify the loads induced by these vortices on immersed bodies. This force partitioning method (and closely related variations of it) has been used to uncover fundamental aspects of aerodynamic force generation in insect flight \citep{Zhang2015CentripetalInsects}, flow-induced oscillations \citep{Menon2020}, flow over delta wings \citep{Li2020VortexWing} as well as other problems in unsteady aerodynamics \citep{Lee2012VorticityPlate,Martin-Alcantara2015VortexAttack,Menon2021a}.

While the task of quantifying the forces induced on finite wings by individual vortices is significantly more complicated in three-dimensional flow-fields, this paper builds off our previous work and demonstrates an application of the force partitioning method in quantitatively estimating the distinct roles that spanwise and cross-span oriented vorticity play in the generation of lift on finite wings. The focus on distinguishing spanwise from cross-span vorticity contributions is motivated by the importance of the spanwise-oriented vortices in the generation of transverse forces; for pitching wings, these are the leading-edge and trailing edge vortices \citep{Menon2021a}, and for a bluff-body, these are the Karman vortices in the wake \citep{Menon2020}. Of particular relevance to this study is the work of Lee \textit{et al.} \citep{Lee2012VorticityPlate}, which used a similar method to analyze the forces on an impulsively started finite flat-plate wing. Their study focused primarily on the distribution of lift along the span of wings with varying aspect-ratios and angles-of-attack. They also discussed the role of different vortices (such as leading-edge and tip vortices) and highlighted three-dimensional effects, by separating the influence of spanwise and transverse vorticity, on the initiation of lift. In the current  work, we provide further insight into such flows by analyzing not just the transient influence of vortex structures over the wing, but also the spatio-temporal evolution of spanwise and cross-spanwise oriented vorticity in the wing's wake. This is performed during the initial transient as well as stationary state of the flow. Crucially, we also analyze the influence of not just the rotation-dominant but also the strain-dominant regions associated with these vortex structures, an effect that is mostly ignored in prior studies. In fact, we will show that this spatio-temporal evolution of different vorticity components in the wake leads to counter-intuitive physics that has important consequences for lift production.

The main aim of this paper is to report on a novel and previously unreported (to our knowledge) physical mechanism regarding the relative contributions to lift production from spanwise and cross-spanwise oriented vorticity on finite wings at low Reynolds number. A stationary wing configuration is chosen for this study, and contributions to lift are examined across angles-of-attack and wing aspect-ratios. By using the force partitioning method this study highlights the important and distinct (and often-ignored) influence of strain-dominated regions of the flow, as well as their interaction with non-linear vortex tilting/stretching mechanisms, in lift generation on finite wings

\section{Force partitioning method and vortex-induced loads}
\label{sec:fpm}
\newcommand{\fpmkin}{C^{\kappa}_{F_i}}
\newcommand{\fpmvif}{C^{\omega}_{F_i}}
\newcommand{\fpmshr}{C^{\sigma}_{F_i}}
\newcommand{\fpmpot}{C^{\Phi}_{F_i}}
\newcommand{\fpmext}{C^{\Sigma}_{F_i}}
\newcommand{\vol}{{V_f}}
\newcommand{\clvif}{C^{\omega}_L}
\newcommand{\clvifnonspan}{C^{\omega xy}_L}
\newcommand{\clvifspan}{C^{\omega z}_L}
\newcommand{\clvifintegrand}{-2Q\phi_2}

\begin{figure}
  \centerline{\includegraphics[scale=1.0]{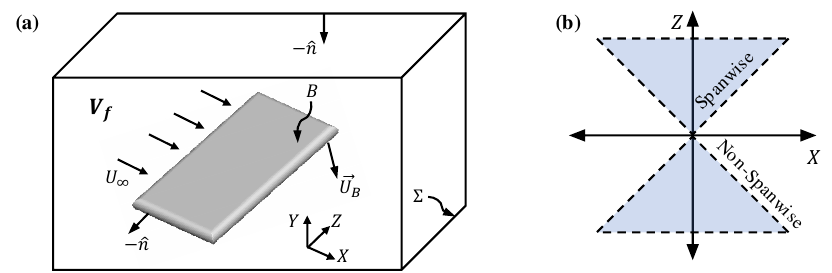}}
  \caption{(a) Schematic of the setup for the force partitioning method (FPM), along with relevant symbols. (b) Schematic showing a projection of the cone separating spanwise and cross-spanwise vorticity in the $X$-$Z$ plane.}
\label{fig:fpm_schematic}
\end{figure}

% \begin{figure}
%   \centerline{\includegraphics[scale=0.8]{plots/fig1.eps}}
%   \caption{(a) Schematic of the setup for the force partitioning method (FPM), along with relevant symbols. (b) Sample snapshot of the $\phi_2$ field around an airfoil.}
% \label{fig:fpm_schematic}
% \end{figure}
A key aspect in this work is the ability to estimate vortex-induced aerodynamic forces due to pressure loads on bodies within fluid flows. This is accomplished by using the force partitioning method described in detail in previous papers by Menon \& Mittal \cite{Menon2020,Menon2021}. Here we provide a concise description of this method, with reference to the schematic in figure \ref{fig:fpm_schematic}. We are interested in quantifying the different mechanisms that contribute to the pressure-induced aerodynamic forces on a body with surface $B$, which is immersed in a flow domain given by $V_f$. The domain is bounded externally by the surface $\Sigma$, and the unit normal pointing outward from the fluid domain into the internal and external boundaries, $B$ and $\Sigma$, is given by $\hat{n}$.

For the partitioning of pressure-induced loads in the $i$-direction on the body, the first step in this method is the calculation of an auxiliary potential field, $\phi_i$, which satisfies the following equation: 
\begin{equation}
  {\nabla}^2 \phi_{i} = 0, \ \ \mathrm{ in } \ \ V_f \ \ \mathrm{ with } \ \
  \boldsymbol{\hat{n}} \cdot \boldsymbol{\nabla} \phi_{i}=
    \begin{cases}
      n_i \;, \; \mathrm{on} \; B \\
      0 \; \;, \; \mathrm{on} \; \Sigma 
    \end{cases}
    \ \ \ \mathrm{ for} \  i=1, 2, 3.
  \label{eq:fpm_potential}
\end{equation}
Here $n_i$ is the component of the unit vector $\hat{n}$ in the $i$-direction. In the current problem, $i=2$ corresponds to the direction of lift forces on the wing and the auxiliary potential field of interest is therefore $\phi_2$. We note that this auxiliary potential is a function only of the instantaneous position and shape of the immersed body and the outer domain boundary. 

The next step is to project the Navier-Stokes momentum equation onto the gradient of the auxiliary potential and integrate over the volume of the fluid domain as follows, 
\begin{equation}
  -\int_{V_f} \boldsymbol{\nabla} p \cdot \boldsymbol{\nabla} \phi_i \  dV =  \int_{V_f} \rho \left[ \frac{\partial \boldsymbol{u}}{\partial t}  +  \boldsymbol{u} \cdot \boldsymbol{\nabla} \boldsymbol{u}
  -  \nu   \nabla^2 \boldsymbol{u}  
  \right]  \cdot \boldsymbol{\nabla} \phi_i \ dV \ \mathrm{ for} \  i=1, 2, 3
  \label{eq:projected_NS}
\end{equation}
By the definition of $\phi_i$ in equation \ref{eq:fpm_potential}, as well as the incompressibility constraint and the divergence theorem, the above equation can be manipulated to express the pressure-induced force on the body as follows,
\begin{equation}
\begin{aligned}
F_i =
  \int_{B}  p \ n_i \  dS  =  
  &- \rho \int_{B} \boldsymbol{\hat{n}}  \cdot \left( \frac{d  \boldsymbol{U}_B}{d t} \phi_i \right) \ dS \\
  &+ \rho \int_{V_f} \boldsymbol{\nabla} \cdot \left( \boldsymbol{u} \cdot \boldsymbol{\nabla}  \boldsymbol{u} \right) \phi_i \ dV \\
  &+ \mu \int_{V_f}  \left( \boldsymbol{\nabla}^2 \boldsymbol{u} \right)  
  \cdot \boldsymbol{\nabla} \phi_i \  dV \\
  &- \rho \int_{\Sigma} \boldsymbol{\hat{n}}  \cdot \left( \frac{d  \boldsymbol{u}}{d t} \phi_i \right) \ dS
\ \  \mathrm{ for} \  i=1 , 2 , 3
\end{aligned}
  \label{eq:FPM}
\end{equation}
where $\boldsymbol{U}_B$ is the local velocity of the immersed surface. 

The four terms on the right-hand-side of equation \ref{eq:FPM} represent distinct components of the force induced by the surface pressure distribution on the body. The first term contains the well-known linear acceleration reaction force (or added mass in potential flow) \citep{Menon2020} as well as the centripetal acceleration reaction, which was shown to be an important lift generation mechanism for insect wings \cite{Zhang2015CentripetalInsects}. In the present case, this unsteady term is zero due to the fact that we are studying force production over static wings. The third and fourth terms represent the effects of viscous diffusion and flow acceleration at the outer boundary of the domain. The latter of the two is generally negligible for large domains \citep{Zhang2015MechanismsInsects}. Further details of these terms and their contributions to aerodynamic forces as well as moments can be found in the work of Zhang \textit{et al.} \cite{Zhang2015CentripetalInsects,Zhang2015MechanismsInsects} and Menon \& Mittal \cite{Menon2020,Menon2021,Menon2021a}.

The second term on the right-hand-side of equation \ref{eq:FPM}, which is the main focus of this study, can be expressed as 
\begin{equation}
 F_i^\omega = \rho \int_{V_f}  \boldsymbol{\nabla} \cdot \left( \boldsymbol{u} \cdot \boldsymbol{\nabla}  \boldsymbol{u} \right) \phi_i \ dV \equiv - \rho \int_{V_f} 2 \ Q \ \phi_i \ dV\ \ \  \mathrm{ for} \  i=1 , 2 , 3 .
\label{eq:VIF}
\end{equation}
Here the quantity $Q$ is defined as $Q = \frac{1}{2}\left(||\boldsymbol{\Omega}||^2 - ||\boldsymbol{S}||^2 \right) $, where $\boldsymbol{\Omega}$ and $\boldsymbol{S}$ are the anti-symmetric and symmetric parts of the velocity-gradient tensor ($\boldsymbol{\nabla u}$) respectively. This scalar quantity signifies the relative strength of local rotation versus strain in the flow-field; positive values of $Q$ correspond to regions where rotation dominates over strain and vice versa. This rotation-dominance condition, $Q>0$, is commonly is used to detect vortices in a flow \citep{Hunt1988EddiesFlows}. 

Given the centrality of $Q$ (and therefore vortices) in this force component, we identify this term as the ``vortex-induced force". This vortex-induced force is found to be a dominant component in a variety of flows including those over flapping wings \citep{Zhang2015CentripetalInsects}, bluff bodies \citep{Menon2020}, pitching foils \citep{Menon2021,Menon2021a} and delta wings \citep{Li2020VortexWing}. As we will show, this is also true for the flow configuration that is a focus of the current paper. It is also useful to note that the volume-integral form of the vortex-induced force term in equation \ref{eq:VIF} allows us to compute the \textit{local} force contribution from any localized region in the flow-field (or flow feature) of interest. This simply involves isolating the volume(s) in the flow-field corresponding to the feature(s) or region(s) of interest, and computing the integral given in equation \ref{eq:VIF} over the appropriate volume(s). This partitioning of the influence of different spatial regions in the flow has been used to effectively dissect the physics of force generation in previous studies \cite{Zhang2015CentripetalInsects,Menon2020,Menon2021,Menon2021a}. Finally, we note that equation \ref{eq:VIF} also indicates that rotation-dominant regions ($Q>0$) and strain dominant regions ($Q<0$) induce pressure forces of opposite signs on the immersed control surface. This simple observation will be useful in dissecting the effect of various vortices on the surface loads.  

A specific focus in this work is on isolating the distinct roles that spanwise and cross-spanwise oriented vortices play in pressure-induced lift production over three-dimensional wings. Our interest in the analysis of spanwise vortices in particular stems from the well-established influence of the leading-edge vortex in a wide variety of practical applications. While there are many methods to identify the volumes in the flow-field occupied by vortices aligned in particular directions, we use a simple method based on the direction of the vorticity unit vector at every point in the flow-field. In our coordinate system, $X$ and $Z$ are the streamwise and spanwise directions respectively, and the three components of the vorticity vector, $\boldsymbol{\omega}$, are denoted as $\omega_x$, $\omega_y$ and $\omega_z$. The component of the vorticity unit vector in the spanwise direction is therefore given by,
\begin{equation}
    \eta_z = \frac{\omega_z}{\sqrt{\omega_x^2 + \omega_y^2 + \omega_z^2}}.
    \label{eq:spanwise_unitvec}
\end{equation}
The magnitude of $\eta_z$ in equation \ref{eq:spanwise_unitvec} represents the cosine of the angle between the spanwise direction ($Z$-axis) and the local vorticity vector at any point in the flow-field. Here we identify ``spanwise oriented'' vorticity very simply as regions where this angle is less than $45^{\circ}$, which implies $|\eta_z|>\mathrm{cos}(\pi/4)$. Geometrically, this condition is satisfied at any point in space if the local vorticity vector lies within either of two symmetric cones whose axes are aligned with the positive and negative $Z$-axes (spanwise direction), with each cone having a half-angle of $45^{\circ}$ and its apex at the point of interest. A schematic with the geometry of these cones is presented in figure \ref{fig:fpm_schematic}(b). Vorticity vectors that lie within this cone are oriented more in the spanwise direction than in any other direction. Conversely, the remaining vorticity vectors are designated as cross-spanwise oriented. For the cases analyzed here, we will show using flow visualizations in subsequent sections that this cross-spanwise component primarily consists of streamwise oriented vortex structures. 
This simple method for separating spanwise from cross-spanwise vorticity that is used here is sufficient for the current purposes but other complex flows would likely require more sophisticated methods, including modal decomposition \cite{Taira2017ModalOverview}, to separate the effect of different vortex structures on aerodynamic loads.

We perform this analysis of spanwise and cross-spanwise vorticity within a subdomain of the full computational grid. The subdomain has dimensions $6C \times 3C \times 3C$ in the $X$, $Y$ and $Z$ directions. It is positioned symmetrically with the wing in the $Y$ and $Z$ directions, and includes $4.55$ chord-lengths in the wake of the wing in the streamwise direction. We have verified that this subdomain captures the bulk of the force (about 95\%) induced on the wing by the surrounding flow. In this study, we are specifically interested in the analysis of the lift production over three-dimensional wings. We report our results in terms of the lift coefficient, $C_L = F_L/(\frac{1}{2} \rho U^2_\infty A)$, where $F_L$ is the dimensional form of the lift and $A$ is the area of the wing. We express the lift coefficient associated with the vortex-induced force as $\clvif$, and the spanwise and cross-spanwise vorticity components of this lift coefficient are $\clvifspan$ and $\clvifnonspan$.

\section{Computational method}
\label{sec:comp_method}

\subsection{Flow solver}
\label{sec:flow_solver}
The incompressible flow simulations in this study have been performed using the sharp-interface immersed boundary method of \cite{Mittal2008ABoundaries} and \cite{Seo2011AOscillations}. This method is particularly well-suited to flows around complex geometries as it allows us the use of a simple non-conformal Cartesian grid to simulate a variety of different shapes and motions of the immersed body. Further, the ability to preserve the sharp-interface around the immersed boundary ensures very accurate computations of surface quantities. The Navier-Stokes equations are solved using a fractional-step method. Spatial derivatives are discretized using second-order central differences in space, and time-stepping is achieved using the second-order Adams-Bashforth method. The pressure Poisson equation is solved using a geometric multigrid method. This code has been extensively validated in previous studies for a variety of three-dimensional stationary and moving boundary problems \citep{Ghias2007,Mittal2008ABoundaries,Seo2011AOscillations}, where its ability to maintain local (near the immersed body) as well as global second-order accuracy has been demonstrated. Further, the accuracy of surface measurements has been established for a wide variety of stationary as well as moving boundary problems in these studies.

\subsection{Problem setup}
\label{sec:problem_setup}

\begin{figure}
  \centerline{\includegraphics[scale=1.0]{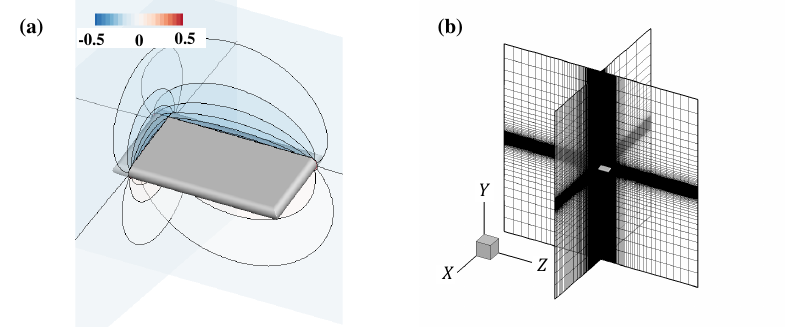}}
  \caption{(a) Geometry of the $AR = 2$ rectangular wing along with spanwise and streamwise slices showing contours of $\phi_2$. (b) Computational domain and the Cartesian computational grid.}
\label{fig:comp_setup}
\end{figure}
In this paper, we analyze the three-dimensional incompressible flow over flat-plate wings with rectangular planforms. Four rectangular wing configurations consisting of span-to-chord aspect-ratios (which we denote as $AR$) of $2:1$ and $3:1$ and angles-of-attack with respect to the freestream flow of $\alpha = 15^{\circ}$ and $\alpha = 25^{\circ}$, are studied. These angles-of-attack are selected because previous work on the low-Reynolds number aerodynamics flows (albeit two-dimensional) over static wings \citep{Menon2020a} has indicated significantly different vortex dynamics in the two cases, with the latter being close to the stall angle at this Reynolds number. In all four cases, the cross section of the wing is a flat plate with a thickness of $15\%$ of the chord. Figure \ref{fig:comp_setup}(a) show a schematic of a case with $AR = 2$ and $\alpha = 15^{\circ}$, with the contours showing the $\phi_2$ field (see equation \ref{eq:fpm_potential}). 

We denote the chord-length of the wing as $C$, and the wing is placed in a uniform incoming freestream flow which has velocity $U_{\infty}$. The chord-based Reynolds number of the flow is $Re = U_{\infty} C /\nu = 1000$. The flow over both wings is simulated in a large computational domain that has dimensions $18C \times 20C \times 20C$ in the streamwise, spanwise and vertical direction. The domain is discretized using a Cartesian grid which is isotropic around the wing and is expanded away from the wing. The grid consists of $192 \times 128 \times 256$ grid cells in the $X$, $Y$ and $Z$ directions. This corresponds to a resolution of approximately 72 points across the chord and 144 points across the span, and at least 5 grid points across the boundary layer (measured at the leading edge where it has least thickness). We have demonstrated in previous work \citep{Menon2019} that this resolution results in well-resolved vortices and shear layers in the low Reynolds number flows studied here. A schematic of the computational domain and the Cartesian grid is shown in figure \ref{fig:comp_setup}(b). A Dirichlet velocity boundary condition is specified at the upstream (inlet) boundary of the domain, and homogeneous Neumann boundary conditions are specified at all other boundaries. The computational cost of each case simulated here is between approximately 60,000 to 220,000 CPU hours on Intel Cascade Lake 6248R cores.
 
\section{Results}
\label{sec:results}

% \subsection{Rectangular wing}
% \label{sec:results_rect}
% \subsubsection{Vortex-induced lift: Contribution of streamwise and spanwise vorticity}
% \subsubsection{Streamwise vorticity near the wing-tip}
% \subsubsection{Streamwise and spanwise vorticity contributions in the wake}

\begin{figure}
  \centerline{\includegraphics[scale=1.0]{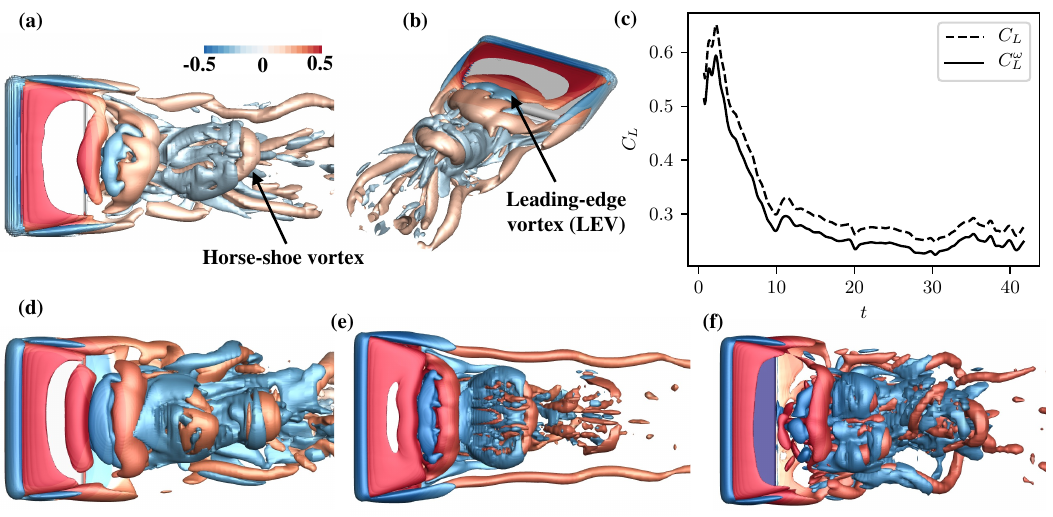}}
  \caption{Vortex-induced lift and flow snapshots showing iso-surfaces of $Q$ at levels [-2.0,2.0], coloured by $\clvifintegrand$ for the four cases studied. (a)-(b) $AR=2$, $\alpha = 15^{\circ}$. (c) Comparison of total lift coefficient ($C_L$) with vortex-induced lift ($\clvif$) for $AR=2$, $\alpha = 15^{\circ}$. (d) $AR=2$, $\alpha = 25^{\circ}$. (e) $AR=3$, $\alpha = 15^{\circ}$. (f) $AR=3$, $\alpha = 25^{\circ}$.}
\label{fig:vif_rect}
\end{figure}
We begin our analysis with a discussion of the vortex-induced lift on the rectangular wings described in section \ref{sec:problem_setup} above. Figure \ref{fig:vif_rect} shows snapshots of the instantaneous flow field for all four cases studied here, visualized using iso-surfaces of $Q$ (at levels [-2.0,2.0]). We see a variety of vortex structures present in the vicinity of the wing as well as in the wake, including both rotation-dominant ($Q>0$) and strain-dominant ($Q<0$) regions of flow. These $Q$ iso-surfaces are colored by the integrand of the vortex-induced force, $\clvifintegrand$, which indicates their local contributions to lift production. For the case with $AR=2$ and $\alpha = 15^{\circ}$ in figures \ref{fig:vif_rect}(a)-(b) we highlight the generation and shedding of a vortex caused by the roll-up of the leading-edge shear layer over the suction surface of the wing as well as prominent tip-vortices. We refer to the former as the leading-edge vortex (LEV). We see that as the LEV sheds into the near-wake, its structure gets slightly deformed from its initial span-wise orientation. Farther downstream, the effect of vortex tilting and stretching transform it into a horseshoe-type structure and leads to the generation of several streamwise-oriented vortices. We see similar dynamics in the other cases with different aspect-ratios and angles-of-attack, with the roll-up and deformation of the leading-edge shear layer to create complex vortical flows in the wake consisting of both streamwise and spanwise vortex structures. It is interesting to note the similarity in vortex dynamics between the two cases with $\alpha = 15^{\circ}$, in figures \ref{fig:vif_rect}(a)-(b) and \ref{fig:vif_rect}(e), which exhibit more coherent vortex structures compared to the two cases at $\alpha = 25^{\circ}$ in in figures \ref{fig:vif_rect}(d) and \ref{fig:vif_rect}(f). We will show in the subsequent discussion that these vortex dynamics and the deformation of the spanwise oriented LEV has important consequences on lift production.

For the case with $AR=2$ and $\alpha = 15^{\circ}$, figure \ref{fig:vif_rect}(c) compares the total lift coefficient on the wing ($C_L$) to the vortex-induced contribution ($\clvif$). It is clear that the pressure load due to vortex-induced effects accounts for the bulk of the lift production on the wing (about 90\%), as is expected. This is true for all the cases studied here (although not shown for brevity). The small difference between the total lift and vortex-induced lift is primarily due to the pressure force induced by viscous momentum diffusion (third term in equation \ref{eq:FPM}), which accounts for 10\%. Viscous shear produces a very small and negative contribution. We do not focus on these viscous effects in this work.

\subsection{Vortex-induced lift: Contribution of spanwise and cross-spanwise vorticity}
\label{sec:strm_span_rect}

\begin{figure}
  \centerline{\includegraphics[scale=1.0]{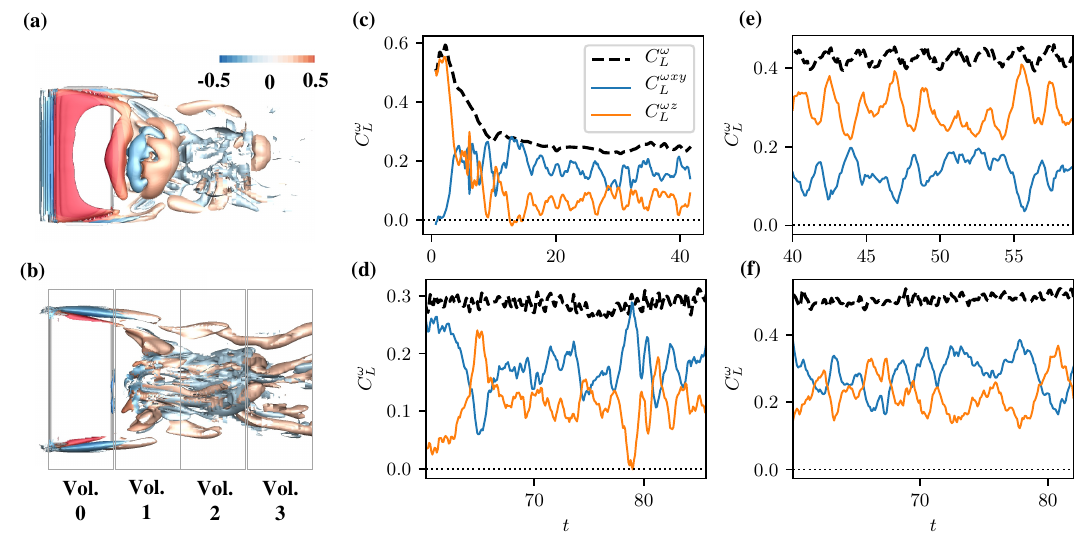}}
  \caption{(a)-(b) Top-view of $Q$ iso-surfaces (levels [-2.0,2.0]) colored by $\clvifintegrand$ for the $AR=2$, $\alpha=15^{\circ}$ wing. Spanwise vorticity structures are shown in (a), by masking $|\eta_z| \leq \mathrm{cos}(\pi/4)$ regions; cross-spanwise vorticity structures are shown in (b), by masking $|\eta_z|>\mathrm{cos}(\pi/4)$ regions. Schematic of distinct cuboidal integration volumes at different downstream distances from the wing are shown in (b). Note that the schematic does not show full extent of integration region in the wake. (c)-(f) Comparison of total vortex-induced lift coefficient ($\clvif$) with contributions from spanwise ($\clvifspan$) and cross-spanwise ($\clvifnonspan$) vorticity for the wings with: (c) $AR=2$, $\alpha=15^{\circ}$; (d) $AR=2$, $\alpha=25^{\circ}$; (e) $AR=3$, $\alpha=15^{\circ}$; (f) $AR=3$, $\alpha=25^{\circ}$. Note (d)-(f) show only stationary state.}
\label{fig:vortdir_rect}
\end{figure}

We now examine the roles that different vortex structures (and their associated strain-fields) play in the production of lift on the wing. As mentioned in section \ref{sec:fpm}, of particular focus in this work is the separate contributions of spanwise and cross-spanwise oriented vorticity in the flow. These contributions are isolated in a very simple manner using the angle of the vorticity unit vector, defined as in equation \ref{eq:spanwise_unitvec}, computed at every point in the domain. Figures \ref{fig:vortdir_rect}(a) and \ref{fig:vortdir_rect}(b) illustrate the flow features that result from this segmentation for one case with $AR=2$ and $\alpha = 15^{\circ}$. The spanwise and cross-spanwise oriented vorticity is visualized by using $Q$ iso-surfaces (levels [-2.0,2.0]) colored by $\clvifintegrand$, and masking the regions that correspond to $\eta_z < 0.7$ and $\eta_z > 0.7$ respectively. 

Figure \ref{fig:vortdir_rect}(a) shows that the spanwise-oriented vorticity, $\eta_z>0.7$, mainly isolates the LEV and shear-layer over the surface of the wing, along with the shedding of the LEV in the near-wake. Farther downstream, there are smaller regions of spanwise vorticity that result from vortex interactions, breakdown, etc. On the other hand, the regions of $\eta_z<0.7$ shown in figure \ref{fig:vortdir_rect}(b) correspond to structures that are largely streamwise oriented, consisting of the tip vortices (although figure \ref{fig:vortdir_rect}(a) shows some spanwise vorticity near the tips, this has a small effect) and a dense region of vortex structures in the mid-span portion of the near-wake. These streamwise structures are a result of vortex-tilting and stretching due to shear in the flow, and include the ``legs'' of the horseshoe vortex that originate from the LEV. Overall, this simple segmentation of rotation-dominated and strain-dominated structures based on the orientation of the vorticity unit vector does a reasonable job of isolating important features of the flow -- the shear-layer, LEV, tip vortices, as well as features that result from the shedding, deformation and tilting/stretching of these structures. 

% \begin{figure}
%   \centerline{\includegraphics[scale=1.0]{plots/options/vortdir_rect_1.pdf}}
%   \caption{\textcolor{red}{OPTION 1}}
% \label{fig:vortdir_rect_1}
% \end{figure}

% \begin{figure}
%   \centerline{\includegraphics[scale=1.0]{plots/options/vortdir_rect_2.pdf}}
%   \caption{\textcolor{red}{OPTION 2}}
% \label{fig:vortdir_rect_1}
% \end{figure}

% \begin{figure}
%   \centerline{\includegraphics[scale=1.0]{plots/options/vortdir_rect_3.pdf}}
%   \caption{\textcolor{red}{OPTION 3}}
% \label{fig:vortdir_rect_1}
% \end{figure}

% \begin{figure}
%   \centerline{\includegraphics[scale=1.0]{plots/options/vortdir_rect_4.pdf}}
%   \caption{\textcolor{red}{OPTION 4}}
% \label{fig:vortdir_rect_1}
% \end{figure}

We can now use equation \ref{eq:VIF}, along with the segmentation described above, to quantify the contributions of the spanwise and cross-spanwise vorticity-containing regions to the total vortex-induced lift. Figures \ref{fig:vortdir_rect}(c)-(f) show this comparison as a function of time for the four cases analyzed in this study, from which several interesting observations can be made. For the case with $AR = 2$ and $\alpha=15^{\circ}$, figure \ref{fig:vortdir_rect}(c) shows that early in the simulation, as the flow develops over the wing, the total vortex-induced lift ($\clvif$) goes through a large-magnitude transient phase. This large lift peak is caused by the formation and eventual shedding of the spanwise oriented starting vortex from the trailing edge and dynamic stall vortex from the leading edge. Soon after this transient, we observe a large drop in the spanwise lift contribution, along with a growth in the cross-spanwise contribution. In fact, the spanwise vorticity eventually has a lower overall contribution to the total lift than the cross-spanwise vorticity. After the initial transient, the contribution of spanwise vorticity to the total lift and vortex-induced lift are approximately 26\% and 29\% respectively, while the cross-span vorticity contribute approximately 60\% of the total lift and 66\% of the vortex-induced lift (the remaining 5\% of vortex-induced lift is accounted for by the flow outside the analysis sub-domain described in section \ref{sec:fpm}). For the case with $AR = 2$ and $\alpha = 25^\circ$, figure \ref{fig:vortdir_rect}(d) shows that the spanwise vorticity contributes approximately 29\% of the total lift and 39\% of the vortex-induced lift, while the cross-span vorticity dominates in this case too, contributing approximately 47\% of the total lift and 63\% of the vortex-induced lift. Similar, spanwise vorticity accounts for a lower share of the lift in the case with $AR = 3$ and $\alpha = 25^\circ$ shown in figure \ref{fig:vortdir_rect}(f), where spanwise vorticity contributes 34\% of the total lift and 44\% of the vortex-induced lift compared to 43\% of total lift and 55\% of vortex-induced lift coming from cross-span vorticity. Spanwise vorticity dominates lift production in only one case studied here, namely $AR = 3$ and $\alpha = 15^\circ$ shown in figure \ref{fig:vortdir_rect}(e). In this case, spanwise vorticity induces 56\% of the total lift and 69\% of the vortex-induced lift whereas cross-span contributes approximately 25\% of the total lift and 31\% of the vortex-induced lift. The smaller contribution of spanwise vorticity than cross-spanwise vorticity to lift induced on the wing in three out of four cases is a surprising and counter-intuitive result given our conventional understanding that the lift over a wing would be dominated by spanwise oriented vortex structures. Another interesting aspect of this comparison is that the growth and subsequent fluctuations of the lift induced by cross-spanwise vorticity mirrors the corresponding fluctuations in spanwise vorticity for all the cases. This hints at the role of vortex tilting in the creation of cross-spanwise vorticity from spanwise vorticity, which is initially the dominant component according to figure \ref{fig:vortdir_rect}(c). In the rest of this paper, we further dissect these lift components into spatial regions as well as rotation-dominated and strain-dominated contributions to uncover additional insights and explain the counter-intuitive dominance of cross-spanwise vorticity in lift production.

\subsection{Lift induced by spanwise and cross-spanwise vorticity in the wake}

In order to better understand the relative contributions of cross-spanwise and spanwise vorticity to the total vortex-induced lift, we now analyze their local contributions within different spatial regions over the wing as well as in the wake of the wing. Our interest in analyzing contributions in the wing's wake is motivated by the observation in section \ref{sec:strm_span_rect} and figure \ref{fig:vortdir_rect} that the growth of cross-spanwise vorticity-induced lift is driven by the shedding, tilting, and stretching of spanwise vorticity in the wake. This is evident from the emergence of significant cross-spanwise vorticity in the wake of the wing, as shown in figure \ref{fig:vortdir_rect}(b), as well as the correspondence between the lift induced by cross-spanwise and spanwise vorticity which follows the initial dominance of the spanwise component in figure \ref{fig:vortdir_rect}(c). Therefore in this section, we analyze this behavior by quantifying the local lift induced by these structures at their inception over the wing and also at increasing downstream distances from the wing as they evolve with the flow in the wake.

The lift induced by vortex structures corresponding to spanwise and cross-spanwise vorticity within distinct regions over the wing and in the wake is estimated by dividing the analysis domain into cuboids that enclose the different regions of interest. The $X$ and $Z$ extent of each of these cuboids is shown schematically in figure \ref{fig:vortdir_rect}(b), where the vertical lines are planes normal to the streamwise ($X$) direction. These planes serve as a simple way to demarcate regions of the flow at increasing downstream distances from the wing. In this analysis, each cuboidal region has a length of 1 chord-length in the streamwise direction, and they span the size of the analysis subdomain (see last paragraph of section \ref{sec:fpm}) in the other two directions. The cuboidal region of interest over the wing is placed so that it is centered at mid-chord, i.e. its upstream and downstream ends are close to the leading and trailing edges respectively. The lift induced by spanwise and cross-spanwise vorticity within each of these regions is then quantified by performing the volume integral in equation \ref{eq:VIF} within each of these regions separately.

\begin{figure}
  \centerline{\includegraphics[scale=1.0]{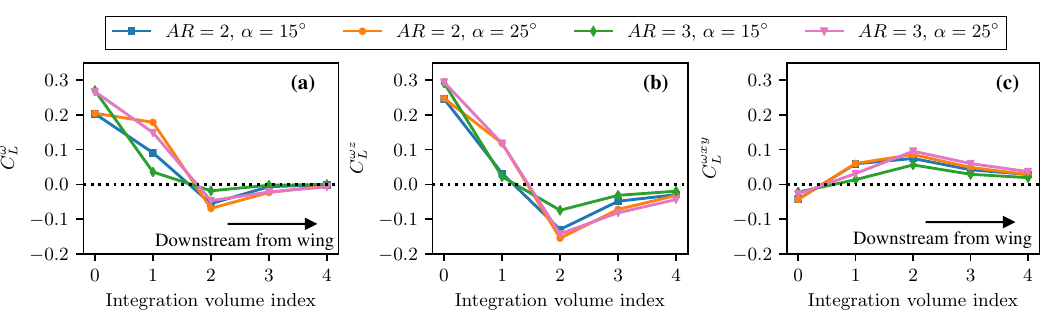}}
  \caption{Time-averaged vortex-induced lift coefficient on the four rectangular wings as a function of downstream distance from the mid-chord. (a) Lift due to all vortex structures ($\clvif$). (b) Lift due to spanwise vorticity containing structures ($\clvifspan$). (c) Lift due to cross-spanwise vorticity containing structures ($\clvifnonspan$).}
\label{fig:vif_x_rect}
\end{figure}

In figure \ref{fig:vif_x_rect} we plot time-averages of the vortex-induced lift coefficient ($\clvif$), as well as spanwise ($\clvifspan$) and cross-spanwise ($\clvifnonspan$) vorticity contributions, within each of these regions as a function of downstream distance from the wing. We first point out that, for all four cases analyzed, figure \ref{fig:vif_x_rect}(a) shows that the largest portion of lift is indeed induced within the region over the wing (downstream distance equals 0). Moreover, figure \ref{fig:vif_x_rect}(b) shows that this lift production over the wing is dominated by spanwise oriented vorticity. As shown earlier in figure \ref{fig:vortdir_rect}(a), the spanwise oriented vorticity over the wing largely corresponds to the LEV and leading-edge shear-layer, which we expect to have a significant contribution to lift. It is interesting to note that the lift induced by spanwise-oriented vorticity in this region seems to be largely determined by the aspect-ratio for the cases studied here, i.e. the wings with the same aspect-ratio generate very similar lift from spanwise-oriented vorticity over the wing. Figure \ref{fig:vif_x_rect}(c) also shows that the cross-spanwise vorticity in this region, which we saw earlier in figure \ref{fig:vortdir_rect}(b) mainly consists of the tip vortices, has a negative effect on lift. This is expected, and is in line our understanding of classical (inviscid) aerodynamics and the downwash-induced effect of tip vortices on finite wings. Moreover, this effect is diminished for the wings with larger aspect ratio, as expected. Therefore on the whole, the lift production mechanisms in the region over the wing align with expectations from theory and prior work for all cases analyzed.

As we move into the wake, we see that the total vortex-induced lift in figure \ref{fig:vif_x_rect}(a) displays an unexpected non-monotonic trend with downstream distance. It first begins to reduce in magnitude within the near-wake compared to the region over the wing, with the different aspect-ratios and angles-of-attack showing different rates of reduction. Farther downstream at about 2 chord-lengths from mid-chord (or 1.5 chord-lengths from the trailing-edge) the vortex-induced lift in fact has a negative value for all four cases. The negative lift has magnitude approximately 27\% and 33\% of the peak value over the wing for the case with $AR = 2$ at $\alpha = 15^{\circ}$ and $\alpha = 25^{\circ}$ respectively. For the case with $AR = 3$, the magnitude of negative lift was found to be 6\% and 17\% of the peak value for $\alpha = 15^{\circ}$ and $\alpha = 25^{\circ}$ respectively. This negative lift is rather surprising as we expect the vortices to have a monotonically diminishing effect with distance into the wake. 

From figure \ref{fig:vif_x_rect}(b), it is clear that the negative vortex-induced lift in the wake is driven by a precipitous drop in the lift induced by spanwise oriented vorticity, which also turns negative within this region. We see that the cases with higher angle-of-attack produce slightly stronger spanwise vorticity-induced negative lift in the wake. The magnitude of the negative lift induced by spanwise vorticity approximately two chord-lengths downstream is roughly 53\% and 62\% of its peak value over the wing for $AR=2$ at $\alpha = 15^{\circ}$ and $\alpha = 25^{\circ}$ respectively, and this is quite significant. The case with $AR = 3$ at $\alpha = 25^\circ$ also shows a similarly high value (about 49\% of its peak value). In contrast, for the case with $AR = 3$ and $\alpha = 15$, this value is 26\%, which is lower compared to other cases. Concurrent with this effect of the spanwise oriented vorticity, Figure \ref{fig:vif_x_rect}(c) shows a \emph{growth} in the lift induced by cross-spanwise vorticity. This growth is however  not large enough to offset the negative lift induced by the spanwise structures, thereby leading to net negative lift from vorticity in the wake. For all four cases, the total vortex-induced lift as well as the contributions from spanwise and cross-spanwise oriented vorticity peak at approximately 2 chord-lengths downstream, and subsequently approach zero as we move farther downstream due to their increasing distance from the wing. This non-monotonic trend in vortex-induced lift as we move into the wake of the wing, and particularly the fact that spanwise vorticity in the wake induces negative lift on the wing while cross-spanwise wake-vorticity induces positive lift, is an intriguing result that is subjected to further analysis here. 

\subsection{The role of rotation-dominated and strain-dominated regions in lift generation}
It has been shown in previous work that although much of the focus in the literature concerning aerodynamics has been on the role that vortices play in force production, vortices are also associated with regions of strain that in fact, have a significant dynamical effect \citep{Menon2021}. In the context of this study, we hypothesize that the drop in lift due to spanwise structures and the subsequent negative lift might be a result of the development and growth of vortex-induced strain associated with these vortices after they shed into the wake of the wing. The strain-dominated regions, which correspond to $Q<0$, would induce negative lift according to equation \ref{eq:VIF}. Furthermore, it can be shown using simple vortex models that vortex cores, i.e. regions of $Q>0$, are generally associated with higher vorticity magnitudes whereas the regions of strain around vortices are larger diffused region with lower magnitudes of vorticity. As a result, these different regions would be affected to different degrees by vortex stretching and tilting due to the non-linear nature of these mechanisms, and can therefore have different effects on the lift production in the wake.

To investigate the effect of rotation and strain further, we decompose the vortex-induced lift associated with spanwise and cross-spanwise vorticity into their corresponding rotation-dominant and strain-dominant regions. This is done by first identifying spanwise and cross-spanwise oriented structures using the vorticity unit vector as before, and then further decomposing each of these into regions of $Q>0$, corresponding to vortex cores, and $Q<0$, corresponding to strain-dominated flow. These 4 separate components of the flow-field are then used to estimate their contributions to lift within each of the integration regions described earlier using equation \ref{eq:VIF}.

\begin{figure}
  \centerline{\includegraphics[scale=0.9]{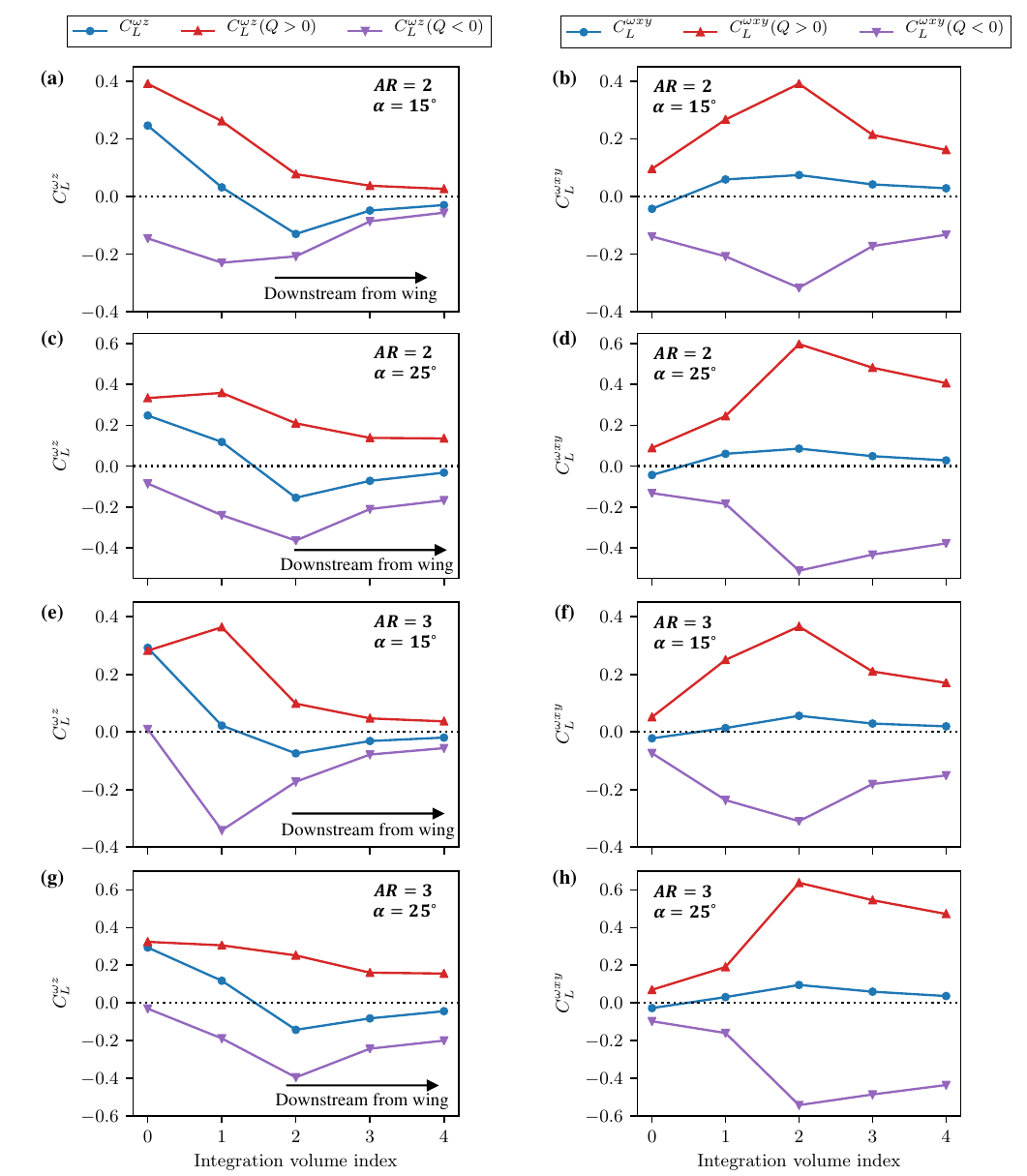}}
  \caption{Time-averaged vortex-induced lift coefficient on the rectangular wing as a function of downstream distance from the mid-chord due to vortex cores ($Q>0$) and strain regions ($Q<0$) corresponding to spanwise vorticity containing structures ($\clvifspan$; left panel) and cross-spanwise vorticity containing structures ($\clvifnonspan$; right panel). (a)-(b) $AR = 2$, $\alpha = 15^{\circ}$. (c)-(d) $AR = 2$, $\alpha = 25^{\circ}$. (e)-(f) $AR = 3$, $\alpha = 15^{\circ}$. (g)-(h) $AR = 3$, $\alpha = 25^{\circ}$.}
\label{fig:vif_x_q_rect}
\end{figure}
In figures \ref{fig:vif_x_q_rect}(a), \ref{fig:vif_x_q_rect}(c), \ref{fig:vif_x_q_rect}(e) and \ref{fig:vif_x_q_rect}(g) we show the time-averaged vortex-induced lift due to the rotation-dominated and strain-dominated regions associated with spanwise oriented vorticity for the four cases analyzed. We first note that within the region around the wing (integration volume index $0$ in figure \ref{fig:vif_x_q_rect}), the positive lift due to rotation-dominated regions of spanwise vorticity, shown in red in figure \ref{fig:vif_x_q_rect}, are the dominant lift producing mechanism for all cases. As discussed before, this corresponds to the effect of the LEV. Conversely, from figures \ref{fig:vif_x_q_rect}(b), \ref{fig:vif_x_q_rect}(d), \ref{fig:vif_x_q_rect}(f) and \ref{fig:vif_x_q_rect}(h), which show the corresponding rotation and strain-dominated contributions from cross-spanwise oriented vorticity, we see that the cross-spanwise vorticity in the vicinity of the wing (i.e. primarily the tip vortices) produces negative total lift, as highlighted earlier. By decomposing the effect of cross-spanwise rotation and strain, we see that the core of the tip vortices ($Q>0$) in fact produces positive lift, but the total effect of the tip vortices is negative due to the strain regions ($Q<0$). We point out that this is in fact consistent with classical finite-wing-theory where the induced velocity due to tip vortices is implicated in lift reduction. In this case, the regions of strain around the tip vortex core correspond to the induced velocity. 

As we move downstream into the wake, the left panel of figure \ref{fig:vif_x_q_rect} indicates an overall drop in the magnitude of positive lift induced by spanwise vortex cores ($Q>0$) 2-3 chord-lengths downstream of the wing for all cases analyzed. This is accompanied by an initial increase in the magnitude of negative lift due to spanwise strain-dominated ($Q<0$) regions (followed by a subsequent decay in the far-wake). Although the rate of this increase varies for different aspect-ratios and angles-of-attack analyzed, the case with $AR=3$ and $\alpha=15^{\circ}$ behaves quite differently from the other three cases. It is the only case with a significant increase in lift induced by spanwise oriented vortex cores in the near-wake. This is likely due to the persistence of coherent spanwise-oriented vortex structures in the near-wake as seen in figure \ref{fig:vif_rect}(e). The tip vortices are also seen to be relatively undisturbed in this case, which suggests there is less interaction and deformation of the spanwise vorticity shed from the wing owing to the stabilizing nature of the larger aspect-ratio and smaller angle-of-attack. Nevertheless, there is a larger increase in negative lift induced by spanwise strain-dominated flow in this case too (as in the other three cases). It is also interesting to note that for the wings at the higher angle-of-attack of $\alpha=25^{\circ}$, the drop in lift induced by spanwise vortex cores is more gradual than at the lower angle of attack. Moreover, the increasing negative lift due to the spanwise oriented strain-dominated vorticity peaks at approximately 2 chord-lengths downstream of the wing for $\alpha=25^{\circ}$, while it peaks closer to the wing at approximately one chord-length for the cases with lower angle-of-attack. Overall for all cases, this disparity between the strain and rotation-dominated regions of spanwise oriented vorticity results in the production of net negative lift in the wake by this component of vorticity.

The right panel of figure \ref{fig:vif_x_q_rect} shows a behaviour for flow structures containing cross-spanwise oriented vorticity in the wake that is opposite to that of the spanwise oriented structures. For all four cases, there is a large increase in the magnitude of positive lift induced by cross-spanwise vortex cores ($Q>0$), and this positive lift peaks two chord-lengths downstream of the wing. There is meanwhile a smaller increase in the magnitude of negative lift due to cross-spanwise straining ($Q<0$) regions. It is interesting to note the similarity in trends between the two pairs of cases with the same angle-of-attack, i.e. the cases with $\alpha=15^{\circ}$ in figures \ref{fig:vif_x_q_rect}(b) and \ref{fig:vif_x_q_rect}(f), as well as the cases with $\alpha=25^{\circ}$ in figures \ref{fig:vif_x_q_rect}(d) and \ref{fig:vif_x_q_rect}(h). This is reflective of the similarity in vortex dynamics discussed earlier and seen in figure \ref{fig:vif_rect}. The net effect of the larger increase in lift induced by cross-spanwise oriented vortex cores compared to strain-dominated regions is positive lift induced in the wake by cross-spanwise oriented vorticity. Therefore, we see for all four cases analyzed that the vortex cores and strain-dominated regions associated with spanwise and cross-spanwise oriented vorticity behave very differently as we move into the wake -- resulting in negative lift induced by spanwise vorticity and positive lift induced by cross-span vorticity in the wake.

%The relative significance of the 4 vorticity components (i.e. rotation and strain-dominated regions for spanwise and cross-span vorticity) in this study can be used to analyze such physical mechanisms. In particular, by noting that $\fpmvif \sim \int_{V_f} Q \, \phi_i dV$ and $\phi_i$ is the same for spanwise and cross-span vorticity components within each integration volume, the trends in figure \ref{fig:vif_x_q_rect} can be used as a proxy for the strength of the 4 vorticity components (i.e. rotation and strain-dominated regions for spanwise and cross-span vorticity) in this analysis.
It is important to note here that the trends in lift induced by the rotation and strain-dominated regions of spanwise and cross-span vorticity shown in figure \ref{fig:vif_x_q_rect} can in fact be used as a proxy to analyze the relative strength of the four vorticity components (i.e. rotation and strain-dominated regions for spanwise and cross-span vorticity). This is clear by noting that $\fpmvif \sim \int_{V_f} Q \, \phi_i \, dV$ and $\phi_i$ is the same for spanwise and cross-span vorticity components within each integration volume. Therefore, the different rates of increase/decrease in the lift induced by vortex cores and their associated strain-dominated regions highlights some important flow physics that affects the balance of these vorticity components in the wake.

\begin{figure}
  \centerline{\includegraphics[scale=1.0]{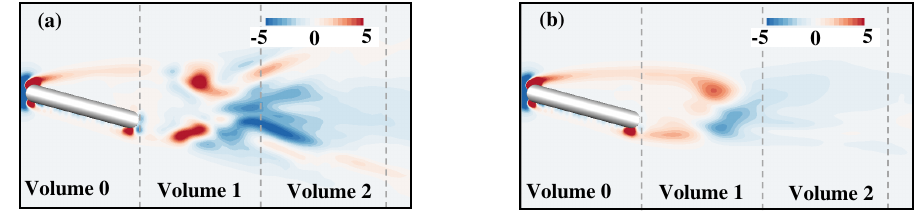}}
  \caption{Contours showing the time-averaged $Q$ field around the $AR=2$, $\alpha = 15^{\circ}$ wing. (a) $X$-$Y$ slice of the domain at mid-span. (b) Slice between mid-span and wing-tip. }
\label{fig:avg_q_rect}
\end{figure}

For one, as pointed out earlier, the larger magnitude of vorticity within vortex cores than their associated strain regions makes vortex cores more susceptible to vortex tilting and stretching due to the non-linearity of these mechanisms. Therefore, we expect the spanwise vortex cores to experience more vortex tilting than their associated straining regions, resulting in more spanwise vortex cores being converted to cross-span vorticity. For all four cases, the left panel of figure \ref{fig:avg_q_rect} reflects this in the more substantial decrease of lift induced by spanwise vortex cores as compared with their associated strain regions as we move into the wake. Furthermore, the growth of vortex-induced strain after the shedding of the LEV and other spanwise oriented vortices in the wake is another factor that augments strain-induced (negative) lift associated with spanwise vorticity in the wake. This strengthening of strain-dominated flow (in addition to less tilting) leads to an increase in the magnitude of negative strain-induced lift as we move into the wake of the wing. To illustrate this latter effect, figure \ref{fig:avg_q_rect} shows snapshots of the time-averaged $Q$ field in the wake of the wing with $AR=2$ and $\alpha=15^{\circ}$. Figure \ref{fig:avg_q_rect}(a) shows a slice of the field in a $X$-$Y$ plane at the mid-span of the wing and figure \ref{fig:avg_q_rect}(b) shows a slice of the field at a location between mid-span and the wing-tip. At both locations, the time-averaged distribution of $Q$ shows the emergence of strong region of strain ($Q<0$ in blue) in the wake of the wing, in regions approximately 1-2 chord-lengths downstream of the wing. This region would therefore enhance the negative lift induced in the wake, as per equation \ref{eq:VIF}, and its location in fact agrees with the peak in negative lift due to strain-dominated spanwise oriented vorticity at approximately 1-2 chordlengths downstream of the wing in figure \ref{fig:vif_x_q_rect}(a). This therefore contributes to a net negative lift induced by spanwise vorticity in the wake.

In the case of the cross-spanwise vorticity in the wake of the wing, the larger increase in the strength of vortex cores compared to strain-dominated regions seen in figure \ref{fig:avg_q_rect}(b) can be explained by the preferential tilting of spanwise oriented vortex cores, as discussed above, to create cross-spanwise vortex cores (streamwise and horse-shoe vortices), as well as the preferential stretching of the streamwise vortex cores. This therefore strengthens the positive lift induced by vortex cores to a larger degree than than the magnitude of negative lift induced by strain-dominated regions -- thus leading to net positive lift induced by cross-span vorticity in the wake. Hence we see that the relative strengths of the the rotation and strain-dominated flow corresponding to spanwise and cross-span oriented vorticity, which is mediated by non-linear vortex stretching and tilting mechanisms, has interesting consequences for the overall lift induced on the wing.

%Another way to interpret the above behavior is as follows: as shown by Menon and Mittal \cite{Menon2021SignificanceLoads}, $Q$ satisfies the condition that $\int_{V_f} Q dV=0$ within a flow-field occupying volume $V_f$, and therefore negative and positive $Q$ has to balance out in the flow domain. However, this balance of positive and negative $Q$ does not have to be satisfied individually for spanwise and cross-span vorticity containing regions. Hence as shown in figure \ref{fig:vif_x_q_rect}, we find that in the near wake, a larger proportion of the $Q<0$ regions are associated with spanwise-oriented vorticity as compared to cross-span oriented vorticity. The negative lift associated with $Q<0$ regions consequently reduces the total lift induced by the spanwise structures and leads to the somewhat counter intuitive result that the spanwise strictures in the near wake generate a net negative lift whereas the streamwise structures in the near wake generate a net positive lift.

\section{Summary and concluding remarks}
\label{sec:summary}

We have demonstrated the use of a rigorous force partitioning method to dissect the physics of lift production in vortex-dominated flows over low aspect-ratio wings. We specifically focus on partitioning the vortex-induced force induced by flow structures containing spanwise and cross-spanwise oriented vorticity on two rectangular wings with aspect ratios of 2:1 and 3:1, and angles-of-attack of $15^{\circ}$ and $25^{\circ}$. The force partitioning method allows us to quantify the pressure-induced force contributions due to these distinct vorticity components, as well as to analyze their influence within different spatial regions of the flow-field. Using these techniques, we were able to demonstrate new insights into the role of spanwise and cross-spanwise oriented vorticity in lift production, which are relevant to such wings.

We showed that the lift production on these low-aspect ratio wing is dominated by the overall contribution of cross-spanwise vorticity for three out of four cases studied. This is counter-intuitive as we expected the spanwise vorticity within the LEV to dominate lift production. To understand this better, we further decomposed these lift-producing components into their contributions within regions of the flow over the wing and in the wake of the wing. Surprisingly, we found that the spanwise component of vorticity in the near-wake region induces a negative lift on the wing for all cases. However, the cross-spanwise vorticity induces positive lift in this region. We explained this difference between the lift induced by spanwise and cross-spanwise vorticity by separately quantifying the influence of vortex cores and their associated regions of strain in the wake for both directional components of vorticity. For all the cases analyzed, we showed that while the positive lift production due to spanwise vortex cores drops rapidly in the wake due to the tilting of spanwise vortex cores, the negative lift induced by the strain-dominated regions of spanwise vorticity increases because of the growth of vortex-induced strain associated with the shed vortices in the wake. This difference in strengths of the spanwise oriented vortex cores and their associated strain regions in the wake of the wing therefore results in a net negative lift induced by spanwise oriented vorticity in the wake. Our analysis of aspect-ratio and angle-of-attack indicated that this growth in spanwise oriented strain-dominated negative lift occurs for all cases studied. Furthermore, only one case -- with the lower angle-of-attack and higher aspect ratio -- showed an initial increase in lift induced by spanwise oriented vortex cores due to the persistence of more stable spanwise vortices. However, this case showed the same trend as the other cases farther downstream, with a drop in lift induced by spanwise vortex cores and net negative lift induced by spanwise oriented vorticity in the wake. On the other hand, positive lift induced by cross-spanwise oriented vortex cores in the wake is enhanced by the tilting of spanwise vortex cores to form cross-span structures and the preferential stretching of these vortex cores. Since their associated strain-dominated regions are affected less by these mechanisms, cross-spanwise vorticity in the wake produces net positive lift on the wing. Hence we showed that the interplay between rotation and strain-dominated regions of spanwise and cross-spanwise oriented vorticity, which is dictated by vortex stretching and tilting, encodes interesting physics that plays a significant role in lift generation. Moreover, this phenomenon is seen to occur for all four cases comprising of different angles-of-attack and wing aspect-ratio that were analyzed.
% On comparing this with the corresponding behavior in the case of the delta wing, we showed that the influence of the wake is significantly diminished due to the suppressed vortex shedding. The suppressed shedding therefore eliminated the negative lift in the wake and results in a significant increase in lift coefficient on the wing.

We expect that a better understanding of the physics underlying vortex-induced lift production will aid the development of physics-based flow control techniques as well as the design of novel wing designs/features for lift enhancement. For instance, several efforts have been made to modify the wing tip of aircrafts to reduce the deleterious effects of the tip-vortices \citep{heyes2005modification,board2007assessment}. Similarly, the analysis of other flow features and physical mechanisms that have an influence on aerodynamic force generation, such as that described here, could help in designing and assessing other such design modifications and flow control techniques.
%For example, the insight provided by this method suggests potential benefits from enhanced streamwise structures in the near wake and mechanisms that exploit the nonlinear vortex dynamics highlighted in this study. The development of this and related ideas is a subject of ongoing and future work. 
While this study has demonstrated the utility of the force partitioning method in uncovering fundamental aspects of the force production in such flows, there are several questions that remain to be answered. For one, how the behavior discussed in this work changes for wings of different Reynolds numbers? It would also be interesting to investigate how the phenomenology reported in this study changes for unsteady wings such as those encountered in bioflight, wing-vortex interaction and wing-gust interaction. Such analyses are interesting future directions for this work.

\begin{acknowledgments}
RM acknowledges support from the Army Research Office (Cooperative Agreement W911NF2120087) as well as NSF CBET-2011619 for this work. This work also benefited from the computational resources at the Advanced Research Computing at Hopkins (ARCH) core facility  (rockfish.jhu.edu), which is supported by the AFOSR DURIP grant FA9550-21-1-0303, and the Extreme Science and Engineering Discovery Environment (XSEDE), which is supported by National Science Foundation grant number ACI-1548562, through allocation number TG-CTS100002.
\end{acknowledgments}

% \appendix

% \section{Appendixes}

%\bibliography{references}

\end{document}